\documentclass[12pt]{article}
\usepackage[dvips]{color,graphicx}
\newcommand{\boldsigma}{\mbox{\boldmath $\sigma$}}
\def\poinc{Poincar\'{e} }
\def\bfq {{\bf q}}
\def\bfK{{\bf K}}
\def\bfL{{\bf L}}
\def\bfk{{\bf k}}
\def\bfp{{\bf p}}  
\def\be{\begin{equation}}
 \def \ee{\end{equation}}
\def\bea{\begin{eqnarray}}
  \def\eea{\end{eqnarray}}

\begin{document}
\title{\hskip9cm NT@UW-02-017\\Light Front Cloudy Bag Model: Nucleon
Electromagnetic Form Factors}

\author{Gerald A. Miller\\
  Department of Physics, University of Washington\\
  Seattle, WA 98195-1560\\
E-mail: miller@phys.washington.edu}

\maketitle

\abstract{The nucleon  is modeled, using light front dynamics,  as a  
 relativistic system of
  three  bound constituent quarks
  emersed in    a cloud of pions. 
  The pionic cloud is important
for understanding low-momentum transfer 
physics, especially the neutron charge
radius, but the quarks are dominant at high values of $Q^2$.
The model achieves a very good description of    existing data
for  the
four electromagnetic elastic form factors.}

\vskip 0.5in

The recent exciting experimental results for the ratio proton elastic
form factors
$G_E/G_M$ (or $QF_2/F_1$)\cite{Jones:1999rz,Gayou:2001qd}
 and the
impending high accuracy data for the neutron
 electric\cite{madey} and magnetic\cite{will} form
factors have re-ignited interest in the venerable goal of understanding
the structure of the nucleon.

The aim of the present paper  is to
present a reasonable, workable model which describes
currently available information  and
makes predictions testable against data taken at higher values of $Q^2$,
or taken for improved accuracy.  The  model  should have enough content
 so that its
ultimate disagreement with experiment  
elucidates some missing piece of physics.
 Poincar\'{e} invariance and chiral symmetry are the principal tools
used to construct the model.

 \poinc invariance is maintained or approximated by using 
light-front dynamics,  in which fields are quantized at a fixed
``time''=$\tau=x^0+x^3\equiv x^+$. The $\tau$-development operator
 is then given by
$P^0-P^3\equiv P^-$. 
 The canonical spatial variable  is 
$x^-=x^0-x^3$, with a  canonical momentum $P^+=P^0+P^3$. The other
coordinates are  ${\bf x}_\perp$ and  ${\bf P}_\perp$.
 The 
 relation between energy and
momentum of a free particle is given by:
$   p^-={p_\perp^2+m^2\over p^+},$
 a relativistic kinetic energy which does not contain
a square root operator. This
allows the  separation of  center of mass
and relative coordinates, 
so that the computed wave functions are frame
independent.
The use of the light front is particularly relevant for calculating form
factors, which are probability amplitudes for an nucleon to absorb a four
momentum $q$ and remain  a nucleon. The initial and final nucleons 
have different total momenta. This means that the final nucleon is 
boosted relative to the initial one, and therefore has  a 
 different wave function.  The light
 front technique allows one 
to use  boosts that are
 independent of interactions.

We are concerned with  the
Dirac $F_1$ and Pauli $F_2$ ($F_2(0)=\kappa$, the anomalous
magnetic moment) form factors.
The 
Sachs form factors are 
$ 
G_E = F_1 - {Q^2/ 4M_N^2} F_2,\;
G_M = F_1 +   F_2.\;$
We use
the  current's ``good" component, $J^+,$ so that 
 $
F_1(Q^2)=
\langle N,\uparrow\left| J^+\right| N,\uparrow\rangle, \;
Q F_2(Q^2) = 
(-2M_N)\langle N,\uparrow\left|
J^+\right| N,\downarrow\rangle,
$
with  nucleon light-cone
 spinors, and in a    frame with $q^+=0$ and
$Q^2=\bfq_\perp^2=q_x^2$.

The model nucleon consists of three relativistically moving, bound constituent
quarks, which are surrounded by a cloud of pions.
The quark aspects\cite{bere76}-
\cite{Miller:2002qb},
will be discussed first.  
The original construction of this 
three-quark model was based on  symmetry 
principles\cite{bere76},\cite{chun91}.
The wave function is anti-symmetric,
a function of relative momenta, independent of reference frame, and 
an eigenstate
of the spin operator. 
 Schlumpf\cite{Schlumpf:ce} applied it to compute
a variety of baryonic properties.
Frank, Jennings \& I \cite{Frank:1995pv} used this model to 
predict  a very strong decrease of $G_E/G_M$ as a function of $Q^2$,
which has now been measured.  Explaining 
the  meaning  of this result
was left  for  a second paper\cite{Miller:2002qb} in which 
 imposing Poincar\'{e} invariance was shown to lead to 
 an analytic result that the ratio
$QF_2/F_1$ is constant for  large $Q^2$ and to a
 violation\cite{ralston} of the
helicity conservation rule.

The  wave function we use is given by 
 \be
\Psi(p_i)=\Phi(M_0^2)
u(p_1) u(p_2) u(p_3)\psi(p_1,p_2,p_3),\quad p_i=\bfp_i  s_i,\tau_i\ee
where $\psi$ is a spin-isospin color amplitude factor,
the $p_i$ are expressed in terms of relative coordinates,  the
$u(p_i)$ are ordinary 
  Dirac spinors, $\Phi$ is a spatial wave function and the repeated indices
$p_i$  are summed over. The specific form of $\psi$ is given in Eq.~(12)
of Ref.~\cite{Miller:2002qb} and earlier in Ref.~\cite{chun91}
The notation is that 
${\bf p}_i =
(p^+_i,{\bf p}_{i\perp})$. The  total
momentum is 
$\bf {P}= {\bf p}_1+{\bf p}_2+ {\bf p}_3.$
The relative coordinates are  
$ \xi={p_1^+/ p_1^++p_2^+},\;
\eta={(p_1^++p_2^+)/ P^+},$
and    
$ {\bf k}_\perp =(1-\xi){\bf p}_{1\perp}-\xi {\bf p}_{2\perp},\;
{\bf K}_\perp =(1-\eta)({\bf p}_{1\perp}
+{\bf p}_{2\perp})-\eta {\bf p}_{3\perp}.$
In  computing  a form factor, we take quark 3 to be the one
struck by the photon.
The value of $1-\eta$ is not changed  ($q^+=0$), so
 only one relative momentum, $\bfK_\perp$ is changed:
${\bfK'}_\perp 
=\bfK_\perp-\eta\bfq_{\perp}.$
We take the  form of the 
spatial wave function from  Schlumpf\cite{Schlumpf:ce}:
 $ 
\Phi(M_0)={N\over (M^2_0+\beta^2)^{\gamma}},
$  
with   $M_0^2$ is the mass-squared operator for a non-interacting system:
\be M_0^2
={K_\perp^2\over \eta(1-\eta)}+
{k_\perp^2+m^2\over \eta\xi(1-\xi)} +{m^2\over 1-\eta}. \ee
Schlumpf's parameters  are
$\beta =0.607\;{\rm GeV}, \; \gamma=3.5,\; m = 0.267\; {\rm GeV}$. The
 value
of 
$\gamma$ was chosen that $ Q^4G_M(Q^2) $ is approximately  constant for
$Q^2>4\; {\rm GeV}^2$ in accord with experimental data. The parameter 
$\beta$ helps govern the values of the perp-momenta allowed by the
wave function $\Phi$ and is closely related to the  rms charge radius,
and $m$ is mainly determined by the magnetic moment of the proton. 
We shall
use different values  when  including the pion cloud.

The calculation of form factors is  simplified by using completeness to
express the wave function
in terms of  light cone
spinors  $u_L(p^+,\bfp ,\lambda)$, which are related to Dirac spinors by a  unitary  Melosh rotation 
 evaluated  in terms of Pauli
spinors: $\vert \lambda_i\rangle,\vert s_i\rangle,$
with $\langle \lambda_i\vert { R}_M^\dagger(\bfp_i)\vert s_i\rangle
\equiv \bar u_L(\bfp_i,\lambda_i)u(\bfp_i, s_i)$.
 Thus the wave function depends on Melosh rotated Pauli spinors:
\be \vert\uparrow \bfp_i\rangle= 
\left[ {m+(1-\eta)M_0+i{\boldsigma}\cdot({\bf n}\times {\bf p}_3)\over
\sqrt{(m+(1-\eta)M_0)^2+p_{3\perp}^2}}\right]
\pmatrix{1\cr 0}\label{spin}\ee
where the quantity in brackets  
is  $
{ R}_M^\dagger(\bfp_3) 
$.
The spin-isospin wave function can then be thought of as
constructed from the non-relativistic quark model, but with the replacement
of Pauli spinors by those of Eq.~(\ref{spin}).  
An  important effect resides in the term $({\bf n}\times {\bf p}_3)$ which
originates from the lower components of the Dirac spinors:
the  orbital angular momentum
$L_z\ne0$\cite{Miller:2002cx}. The term $({\bf n}\times {\bf p}_3)$
is also responsible for the flatness of the ratio $F_2(Q^2)/F_1(Q^2)$.

\begin{figure}
\unitlength.9cm
\begin{picture}(5,5)(-3,4.6)
\includegraphics{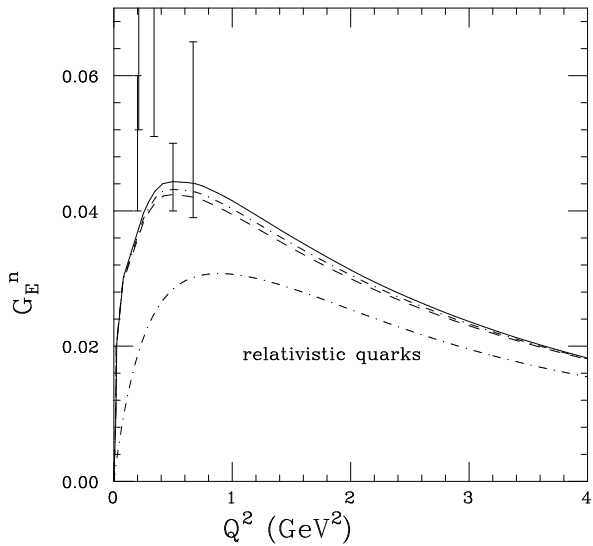}
\end{picture}\vspace{1.7cm}
\caption{\label{figgen}Calculation of $G_E^n.$
  The data
  are from  Ref.~\cite{nrefs}.} 
\end{figure}

We turn now to neutron properties. 
The three-quark model for the 
proton respects charge symmetry, invariance under the 
 interchange of $u$ and $d$ quarks,
so  it  contains a prediction, 
shown in Fig.~\ref{figgen} (compared with data
 from  Ref.~\cite{nrefs})
for neutron form factors. We note that $G_{En}$ would vanish
in the non-relativistic limit, $R_M\to1$, so the deviations from 0 are solely
due to relativistic effects. 
The resulting electric form factor, shown in the 
curve labeled relativistic quarks, is 
very small at low values of $Q^2$, 
  but at larger values of $Q^2$
the prediction is  larger than that of the Galster 
parametrization\cite{galster}. 

The slope of $G_{En}$ is related to the charge radius
as
$G_{En}(Q^2)\to-Q^2R_n^2/6$ with a measured value\cite{nrm}
of $R_n^2=-0.113 \pm 0.005 \;$fm$^2$.
The three-quark model value is 
 $-0.025$ fm$^2$. To understand this  small magnitude
 we express $G_E$
  in terms of $F_{1,2}$ for small values of $Q^2$.
Then 
$ R_n^2=
R_1^2 +    R_F^2, $ 
where the Foldy contribution, 
$R_F^2=6 \kappa_n/4M^2=-0.111\; {\rm fm}^2$ is,  by itself,
in good agreement with
the experimental data. But this does not guarantee success in 
 explaining the charge radius because 
one needs to
include the $Q^2$ dependence of $F_1$ which gives $R_1^2$. In the three-quark
model $R_1^2=+0.086\; {\rm fm}^2$ which nearly cancels the effects of $R_F^2$.
Such a 
 cancellation is a 
natural consequence of including the relativistic
effects of the lower components 
of the quark Dirac spinors\cite{Isgur:1998er}.
Another effect is needed.

\begin{figure}
\unitlength.9cm
\begin{picture}(5,5)(-3.0,8.0)
\includegraphics{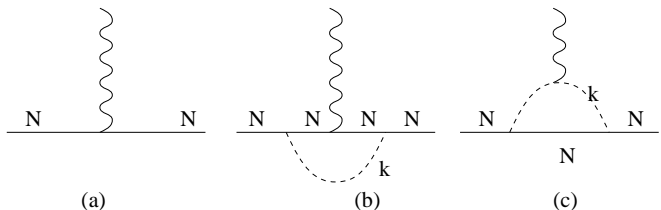}
\end{picture}
\caption{\label{fig:diagrams}Diagrams}
\end{figure}

Sometimes a physical nucleon can be a  bare nucleon emersed in  
 a pion
cloud. An incident photon can interact electromagnetically with
a bare nucleon, Fig.~\ref{fig:diagrams}a,  with a nucleon while 
a pion is present, Fig.~\ref{fig:diagrams}b, or with a charged
pion in flight, Fig.~\ref{fig:diagrams}c.
These effects are especially
pronounced for the neutron $G_E$\cite{cbm}, at small values of $Q^2$,
because the
quark effects are  small.
 The tail of the negatively charged
pion distribution extends far out into
space, causing  $R_n^2$ 
to be  negative. Such  contributions were computed long ago
using  the cloudy bag model\cite{cbm}, which employed
static nucleons.

It is necessary to compute the effects of the pion cloud
in a relativistic manner, to  confront  data taken at large $Q^2$.
This involves   evaluating  the Feynman diagrams of
Fig.~\ref{fig:diagrams}
using photon-nucleon form factors from our
relativistic model, and
using a relativistic $\pi$-nucleon form factor. We define  the resulting model
as the light-front cloudy bag model LFCBM. 
The light-front treatment is implemented by doing the integral over the
virtual pion four-momentum $k^\pm,\bfk_\perp$ 
performing the integral over $k^-$ analytically, re-expressing the
remaining integrals in terms of relative variables ($\alpha=k^+/P^+)$, and
shifting the relative $_\perp$ variable to $\bfL_\perp$
to simplify the numerators.
Thus the Feynman graphs, Fig.~\ref{fig:diagrams}, are represented by a single
$\tau$-ordered diagram. The  use of $J^+$ and the
Yan identity\cite{Chang:qi} $S_F(p)=\sum_s
u(p,s)\overline{u}(p,s)/(p^2-m^2+i\epsilon)+\gamma^+/2p^+$
allows one see that the nucleon current operators appearing in
Fig.\ref{fig:diagrams}b
act between on-mass-shell spinors.

The results  can be stated as
\bea
F_{i\alpha}(Q^2)=Z\left[F_{i\alpha}^{(0)}(Q^2) + F_{ib\alpha}(Q^2)
  +F_{ic\alpha}(Q^2)\right],\label{tot}\eea
where $ i=1,2$ denotes  the Dirac and Pauli  form factors, $\alpha=n,p$ determines
the identity of the nucleon, and $F_{i\alpha}^{(0)}(Q^2)$ are the form factors
computed in the absence of pionic effects.
The wave function renormalization constant
$Z$ is determined from the condition the charge of the proton be unity:
$F_{1p}(Q^2=0)=1.$ Calculating the graph Fig.~\ref{fig:diagrams}b gives 
\bea
F_{1bn}(Q^2)=(2F_{1p}^{(0)}(Q^2)+F_{1n}^{(0)}(Q^2))
\int_N(\alpha^2M^2+L_\perp^2-\alpha^2Q^2/4)
\nonumber\\
+(2F_{2p}^{(0)}(Q^2)
+F_{2n}^{(0)}(Q^2))\int_N(\alpha^2Q^2/2),\label{tb1}\eea
\bea
F_{2bn}(Q^2)=(F_{1p}^{(0)}(Q^2)+F_{1n}^{(0)}(Q^2)/2)\int_N(2M^2\alpha^2)
+(F_{2p}^{(0)}(Q^2)+F_{2n}^{(0)}(Q^2)/2)
\int_N(4\alpha^2M^2+2\mu^2).\label{tb2}\eea
where the integration measure $\int_N$ is given by
\bea
\int_N\equiv{g^2\over2 (2\pi)^3}\int d^2L_\perp {d\alpha\over\alpha}
R({\bfL_\perp^{(+)}}^{\;2},\alpha)
R({\bfL_\perp^{(-)}}^{\;2},\alpha),\eea
$g$ is the $\pi$N coupling constant,
$g^2/4\pi=14\;$,$ \bfL_\perp^{(\pm)}\equiv \bfL_\perp\pm\alpha\bfq_\perp/2$,
$\alpha
D(k_\perp^2,\alpha)\equiv{M^2\alpha^2+k_\perp^2+\mu^2(1-\alpha)}$,
and
$ 
R(k_\perp^2,\alpha)\equiv{F_{\pi N}(k_\perp^2,\alpha)\over
  D(k_\perp^2,\alpha)}. $ The  $\pi$N form factor is
taken as
\bea
F_{\pi N}(k_\perp^2,\alpha)=
\exp{[-(D(k_\perp^2,\alpha)/2(1-\alpha)\Lambda^2)]},\eea
as 
used by Refs.~\cite{Zoller:1991cb} and \cite{Szczurek:gw}, and
satisfies the constraints needed  to maintain charge
conservation\cite{Speth:pz}.
Including the form factor this way uses the   assumption that
the  form factor is an analytic function  of 
$k^-$.
The results (\ref{tb1},\ref{tb2}) show that each term in the nucleon
current operator  contributes to both $F_1$ and $F_2$. 
The evaluation of  graph \ref{fig:diagrams}c yields  
\bea
F_{1cn}(Q^2)=-2F_\pi(Q^2)\int_N{(1-\alpha)\over\alpha}
(\alpha^2M^2+L_\perp^2-(1-\alpha)Q^2/4)\label{tc1}\\
F_{2cn}(Q^2)=-2F_\pi(Q^2)\int_N{(1-\alpha)\over\alpha} (2m^2\alpha(1-\alpha)).
\label{tc2}\eea
The proton form factors can be obtained by simply making the replacements
$n\to p$ in Eqs.~(\ref{tb1},\ref{tb2}) and $-2\to+2$ in
Eqs.~(\ref{tc1},\ref{tc2}). 
\begin{table}
  \centering
  \caption{Different parameter sets, units in terms of  ${\rm fm}$  }
  \vspace{0.1cm}
  \begin{tabular}{|l|rrrrrrr|}\hline
 {\em Set(legend)} & 
 $m$ & $\beta$ & $\Lambda$ & $\gamma$ &-$R^2_n$& $-\mu_n$& $\mu_p$\\ 
\hline
      1 solid & 1.8 &3.65 & 3.1 &4.1&0.111&1.73&2.88\\
      2 dot-dash & 1.7 &3.4& 3.1 &3.9&  0.110&1.79&   2.95\\
      3 dash &1.7&  2.65 & 3.1 & 3.7&   0.109&    1.79          &     2.95\\   
       \hline
      \end{tabular}
      \end{table}

Eqs.~(\ref{tot}-\ref{tc2}) completely specify the form of the 
calculation.
But the LFCBM
requires  four parameters $m,\beta,\gamma,\Lambda$. Including pionic effects
while continuing to use the original values of $m,\beta,\gamma$, would lead to
a satisfactory description of $G_{En}$\cite{Miller:2002cx}, but would cause
other computed observables
 to disagree with experiment.
Thus a new set of parameters is needed. The following        
set of requirements is used to restrict  the parameters. First,
the magnetic moments
of the proton and neutron must agree with measured ones within 10\%.
We also require that 
the computed values of
$G_{Mn}(0.5),\;$$G_{En}(1,1.5),\;$$G_{Mn}(4),\;$
$\mu G_{Ep}/G_{Mp}(5.5),\;$$ G_{Mp}(5.5)$ and 
$G_{Mp}(10)$
agree with the  measured  values well enough so that the average disagreement
is about one error bar. There are many parameter sets that satisfy this
criterion. We show results for three  in Table~1\cite{code} and in the figures.

\begin{figure}
\unitlength1cm 
\begin{picture}(5,5)(-2,4.5)
  \includegraphics{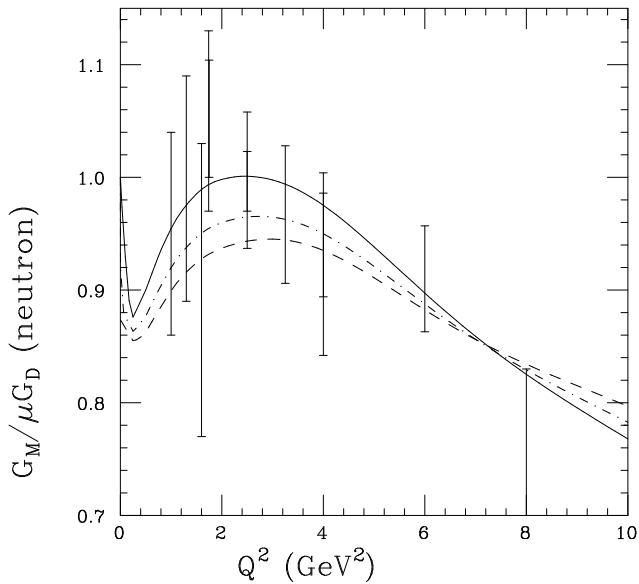}
\end{picture}\vspace{2.3cm}
\caption{\label{gmnodip} $G_{Mn}/\mu G_D$ for the neutron.
  Data 
  are from Ref.~\cite{gmn}. }
\end{figure}

The first application of the LFCBM is to $G_{En}$ and the results
of using the three parameters sets of Table~1 are shown in
Fig.~\ref{figgen}. It is easy to find many parameters which
provide a large pionic effect at small values of $Q^2$. The agreement with
existing data is good, and more higher quality data at larger values of
$Q^2$ is expected\cite{madey}. 
The next step is to compute $G_{Mn}$, which is expressed as
$G_{Mn}/\mu G_D$, where $\mu$ is the computed neutron magnetic
moment and $G_D=(1+Q^2/0.71{\rm GeV}^2)^{-2}$. The
results are shown  in Fig.~\ref{gmnodip}. The  agreement between
the present theory and existing data \cite{gmn}
is excellent, but this will soon
be tested by a new experiment\cite{will}.

\begin{figure}
\unitlength.9cm
\begin{picture}(8,6)(-3,3.50)
\includegraphics{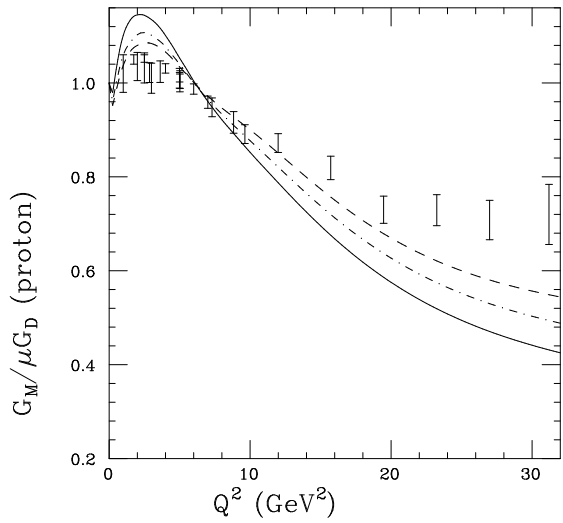}
\end{picture}
\vspace{1cm}
\caption{\label{gmp} $G_{Mp}/\mu G_D$. Data are from Ref.~\cite{gmp} }
\end{figure}

We turn now to the calculation of proton observables.
\begin{figure}
\unitlength1cm 
\begin{picture}(5,5)(-2,5.4)
  \includegraphics{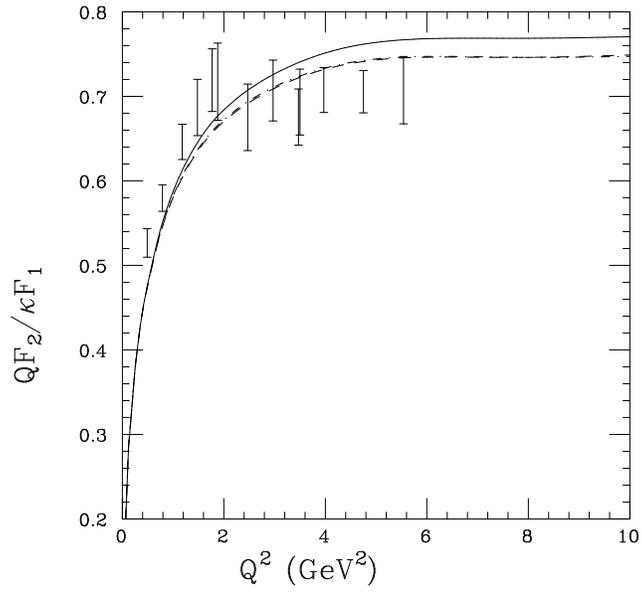}
\end{picture}\vspace{2.8cm}
\caption{\label{f2rat} $QF_2/\kappa F_1$ for proton.
  Data 
  are from  Refs.~\cite{Jones:1999rz,Gayou:2001qd}.}
                                \end{figure}
Fig.~\ref{f2rat} shows that
the measured ratio of Dirac to Pauli form factors is reasonably well-reproduced. This is very similar to our earlier
results\cite{Frank:1995pv,Miller:2002qb}, 
showing that the pion cloud effects are not very important for
this ratio at relatively large values of $Q^2$. 
These ratios  are 
insensitive to the parameter set, and 
the results for sets 2 and 3 overlap.  Thus, as stressed
elsewhere\cite{Miller:2002qb,Miller:2002cx} respecting
symmetries is more important than including 
detailed dynamics in obtaining a constant ratio.
Finally, the  proton magnetic form factor,
 expressed as
$G_{M}/\mu G_D$, where $\mu$ is the computed proton magnetic
moment, is shown in Fig.~\ref{gmp}. 
For this case, set 3 seems to provide
a ``best'' description of the present data\cite{gmp} up to about
$Q^2=20\;{\rm GeV}^2$. For higher values the calculation falls a bit
below the data,
perhaps indicating the need for the effects of perturbative QCD.

These calculations show  that 
the combination of 
\poinc invariance 
and chiral symmetry, as 
realized in the present light front cloudy bag model, is sufficient
to describe the existing experimental data up to about $Q^2=20\;{\rm GeV}^2$.
This is somewhat surprising as the model keeps onlytwo  necessary
effects. 
 The effects of mixing of quark-configurations\cite{Cardarelli:2000tk}, 
 the variation of the quark mass with $Q^2$\cite{qvm}, 
exchange currents\cite{weber} and an intermediate $\Delta$\cite{cbm}
 could enter, but  these seem to have little
 influence within our model. 

Perhaps the strongest feature of the model is
 that it is testable  in upcoming experiments.
For the proton,  $QF_2/F_1$ is predicted to be constant
for values of $Q^2$ up to
 about 20 GeV$^2$. The neutron $G_{En}$,  soon to be measured, is predicted
as is its magnetic form factor  also soon to be measured.

There also  are  implications for other reactions.
Hadron helicity conservation predicts  that
$Q^2F_2/F_1$ is constant\cite{BL80}. 
This is not respected in present data, so
 there is no
need to expect it to hold for a variety of exclusive reactions occurring
at high $Q^2\le 5.5$ GeV$^2$.
Examples include the large spin effects observed in 
 $pp$ elastic scattering\cite{cosbie}
and the
 reaction $\gamma d\to np$\cite{Wijesooriya:yu},
but there are many other possibilities.

\section*{Acknowledgments} This work is partially supported by the U.S. DOE.
I thank R.~Madey for encouraging me to compute $G_E^n$.


\begin{thebibliography}{99}
\bibitem{code} The author will make available the computer program
lfcbm.f to any researcher wishing to choose different parameters.
\end{thebibliography}
\end{document}